\documentclass[english,aps,pre]{revtex4-1}
\usepackage[T1]{fontenc}
\usepackage[latin9]{inputenc}
\usepackage{color}
\usepackage{amsmath}
\usepackage{amssymb}
\usepackage{graphicx}
\usepackage{esint}

\makeatletter
\usepackage{babel}

\usepackage{babel}

\makeatother

\usepackage{babel}
\begin{document}

\title{Epidemic Spreading in Random Rectangular Networks}

\author{Ernesto Estrada$^{1}$, Sandro Meloni$^{2,3}$, Matthew Sheerin$^{1}$,
Yamir Moreno$^{2,3,4}$}

\address{$^{1}$Department of Mathematics \& Statistics, University of Strathclyde,
26 Richmond Street, Glasgow G1 1XH, UK, }

\address{$^{2}$Department of Theoretical Physics, University of Zaragoza,
50018 Zaragoza, Spain, }

\address{$^{3}$Institute for Biocomputation \& Physics of Complex Systems
(BIFI), University of Zaragoza, 50018 Zaragoza, Spain, }

\address{$^{4}$Complex Networks and Systems Lagrange Lab, Institute for Scientific
Interchange, Turin, Italy.}
\begin{abstract}
The use of network theory to model disease propagation on populations
introduces important elements of reality to the classical epidemiological
models. The use of random geometric graphs (RGG) is one of such network
models that allows for the consideration of spatial properties on
disease propagation. In certain real-world scenarios---like in the
analysis of a disease propagating through plants---the shape of the
plots and fields where the host of the disease is located may play
a fundamental role on the propagation dynamics. Here we consider a
generalization of the RGG to account for the variation of the shape
of the plots/fields where the hosts of a disease are allocated. We
consider a disease propagation taking place on the nodes of a random
rectangular graph (RRG) and we consider a lower bound for the epidemic
threshold of a Susceptible-Infected-Susceptible (SIS) or Susceptible-Infected-Recovered
(SIR) model on these networks. Using extensive numerical simulations
and based on our analytical results we conclude that (\textit{ceteris
paribus}) the elongation of the plot/field in which the nodes are
distributed makes the network more resilient to the propagation of
a disease due to the fact that the epidemic threshold increases with
the elongation of the rectangle. These results agree with accumulated
empirical evidence and simulation results about the propagation of
diseases on plants in plots/fields of the same area and different
shapes.

\medskip{}

\end{abstract}
\maketitle

\section{Introduction}

The study of epidemiological models on networks is one of the areas
that has observed a major development in the application of network
theory to real-world problems \citep{Dynamics review}. The discovery
of the fact that networks with fat-tailed degree distributions do
not display an epidemic threshold in the asymptotic limit is a relevant
example of how the connectivity pattern of interacting agents can
dramatically change the course of an epidemic \citep{Epidemics_1,Epidemics_3}.
The use of network theory in epidemiological models provides a way
to incorporate the individual-level heterogeneity necessary for the
mechanistic understanding of the spread of infectious disease. These
characteristics are very attractive for the application of network
epidemiological models in ecology on the different spatial and temporal
scales. 

Although there has been many succesful applications of network theory
to human and animal epidemiology, the situation is a little less developed
for epidemic on plants. Ten years ago Jeger et al.~\citep{plants review}
recognized the relatively low use of network theory for studying plant
diseases. Since then, more theoretical developments have been presented
in the literature. These models include the important description
of the geometric constraints in which the pathogen is spreading as
well as stochasticity and several sources of heterogeneity in the
transmission of infection \cite{Handford,Neri_1,Neri_2,Perez-Reche}. 

In order to consider spatial effects in the transmission of diseases
it is possible to consider spatial networks that treat interactions
as a continuous variable that decays with increasing distance or by
distributing randomly and independently a set of vertices on the Euclidean
plane to represent the relative spatial location of individual hosts
or habitat patches. The second kind of model is based on \textit{random
geometric graphs} (RGGs) \citep*{Penrose,Dall Christense,Bollobas book,Gilbert model},
in which each node is randomly assigned geometric coordinates and
then two nodes are connected if the (Euclidean) distance between them
is smaller than or equal to a certain threshold $r$. Random geometric
graphs have found applications to model populations which are geographically
constrained in a certain region \citep{Spatial connectivity,RGG Sync,Worm Epidemics,RGG spreading,RPG epidemics,RPG cities},
which offer many valuable features over other types of random graphs
\cite{Riley,Opinion Dynamics}. Brooks et al.~\citep{RGG plants}
have used RGGs to model the interactions between the anther smut fungus
and fire pink using temporal data that spans 7 years of field studies.
They have concluded that the use of spatially explicit network models
can yield important insights into how heterogeneous structure promotes
the persistence of species in natural landscapes.

When studying the propagation of diseases in plants there is an important
factor that needs to be taken into account. It is obvious that plants
are not as mobile as humans and animals, thus they reach lower levels
of mixing in a given population due to mobility. The immediate consequence
of this lack of mobility is the fact that the shape of the plot or
field in which the plants are distributed may significantly affect
disease dynamics. In fact, there is both empirical and theoretical
evidence that supports this hypothesis \citep{square plots,square fields_1,plot shapes,square fields_2,rectangular field coconut,plot size_1,computer simulations_1,stochastic model,mathematical model}.
In general, it has been suggested that square plots and fields favored
higher spreading of plant diseases than elongated ones of the same
area \citep{square plots,square fields_1,plot shapes,square fields_2}.
We should make here some remarks about the shape of plots in different
scenarios. First, we should mention the experimental plots for different
crops. In those cases, the size and shape of the plots is controlled
typically to estimate crop yields. Thus, they are typically of almost
perfect square or rectangular shapes (see for instance \cite{plot shapes}).
The second scenario is when crops are cultivated in country fields.
In these cases the sizes and shapes of the fields depends on the geographical
conditions of the region. However, in general these fields can be
grossly approximated as rectangle-like or square-like on the basis
that they are more or less elongated. Such shapes are also though
to facilitate the mechanized work on the fields than more irregular
shapes. Finally, these is a third scenario in which plants are growing
naturally in a given environment. In these cases it is obvious that
the distribution could be quite irregular and acquiring many different
shapes. However, when studying the influence of the shape of these
natural fields on the propagation of an epidemics it is typical to
approximate their shapes to rectangular/square ones, as it is well
illustrated for the case study of the spatial and spatiotemporal pattern
analysis of coconut lethal yellowing in Mozambique \cite{rectangular field coconut}.

It is important to remark that the area of the field also plays a
fundamental role, with larger plots and fields favoring more the spreading
of diseases \citep{plot size_1,stochastic model,mathematical model}.
Also, the orientation of elongated fields may affect the disease propagation
with orientations perpendicular to prevalent winds suppressing epidemic
progression \citep{square fields_1,square fields_2}. All in all,
for plots and field of the same area and orientation there is empirical
and theoretical evidence that elongated shapes decreases the impact
of epidemics on plant populations. It is worth noting that the theoretical
models \citep{computer simulations_1,stochastic model,mathematical model}
used in the previously mentioned studies do not use network theory
as a tool for the study of epidemic spreading.

In this work we consider a generalization of the RGG known as the
\textit{random rectangular graph} (RRG) model which has been recently
introduced \citep{Estrada_Sheerin_1}. In this case, the nodes are
uniformly and independently distributed on a unit rectangle of given
side lengths. Thus, we simulate plots/fields of the same size in which
we can analyze the effect of elongation of epidemic spreading on plants.
When both sides are of the same length we recover the RGG which accounts
for squared shapes. It is worth mentioning that previous models have
considered the variation of the shape for the region where the nodes
are distributed in the RGG. In some of these works more general boundaries,
such as right prisms and fractal regions have been considered in particular
for analyzing the connectivity of the resulting RGGs \citep{Non_square_1,Non_square_2,Non_square_3}.
In this work we consider a Susceptible-Infected-Susceptible
model on the RRGs to describe the propagation of a disease on a plant
population on a field of varied rectangular shapes. We obtain analytical
and simulation results that support the empirical observations and
theoretical evidence about the fact that (\textit{ceteris paribus})
elongated plots/fields decrease significantly the propagation of diseases
on plants. In particular, our results show that the epidemic threshold
is significantly displaced to the right with the elongation of the
rectangle, which indicates that the number of infected plants necessary
to produce an epidemic grows with the rectangle elongation. Finally, we stress that in classical, noninteracting systemsÑeither homogeneous or heterogeneousÑthe resemblance of SIS and SIR 
epidemiological models translates into a strong mathematical symmetry between them that leads to identical expressions for the epidemic thresholds under mean-field approaches (i.e., when neglecting the effects of dynamical correlations). We therefore anticipate that our results will be also valid in a SIR framework, which as a matter of fact could be more relevant for plant diseases.

\section{Random Rectangular Graphs}

Here we consider a population, e.g., plants, represented by the nodes
of a graph for which the edges represent the interaction between the
individuals in the population. Then, our representation consists of
\textit{simple graphs} $G=(V,E)$ defined by a set of $n$ nodes $V$
and a set of $m$ edges $E=\{(u,v)|u,v\in V\}$ between the nodes.
These graphs are unweighted, undirected, with no self-loops (edges
from a node to itself), and no multiple edges. The matrix $A=\left(A_{ij}\right)$,
called the \textit{adjacency matrix} of the graph, has entries 
\[
A_{ij}=\left\{ \begin{array}{ll}
1 & \mbox{if }(i,j)\in E\\
0 & \mbox{otherwise}
\end{array}\right.\qquad\forall i,j\in V.
\]

Once the structure of a network is defined, the adjacency matrix is
not changed during the process of disease propagation to be modeled
on the nodes and edges of that network. That is, the network topology
is static and not changing with time. The degree $k_{i}$ of the node
$i$ is the number of edges incident to it, equivalently $k_{i}=\sum_{j}A_{ij}$.
Let $G=(V,E)$ be a simple connected graph and let $\lambda_{1}>\lambda_{2}\geq\cdots\geq\lambda_{n}$
be the eigenvalues of its adjacency matrix. The eigenvalue $\lambda_{1}$
is known as the principal eigenvalue of the adjacency matrix, also
as the Perron-Frobenius eigenvalue. Below we show that $\lambda_{1}$
is key to determine the conditions of invasion. 

When modeling epidemic disease propagation on plants, Brooks et al.
\citep{RGG plants} have considered the plants as the nodes of a RGG,
in which the $n$ nodes are points uniformly and independently distributed
in the unit square $[0,1]^{2}$ \citep{Penrose}. Then, two points
are connected by an edge if their Euclidean distance is at most $r$,
which is a given fixed number known as the \textit{connection radius}.
This connection radius indicates the maximum distance at which a disease
can be transmitted from one plant to its nearest neighbors (see \citep{RGG plants}).

Here we use an extension of the RGG to consider a rectangle $[0,a]\times[0,b]$
where $a,b\in\mathbb{R},\ a\geq b$. Due to the accumulated evidence
that reveals the importance of the plots/field size on disease propagation
we will keep the area of the rectangle fixed in order to analyze only
the influence of the rectangle elongation on the disease spreading.
Consequently, we will consider unit rectangles of the form $\left[0,a\right]\times\left[0,a^{-1}\right]$.
The rest of the construction of an RRG is similar to that of an RGG.
That is, we distribute uniformly and independently $n$ points in
the unit rectangle $[0,a]\times[0,a^{-1}]$ and then connect two points
by an edge if their Euclidean distance is at most $r$. Obviously,
when $a=1$ the rectangle $[0,a]\times[0,a^{-1}]$ is simply the unit
square $[0,1]^{2}$, which means that the $RRG$ becomes the classical
$RGG$ \citep{Estrada_Sheerin_1}.

\begin{figure}[h]
\begin{centering}
\includegraphics[width=0.5\textwidth]{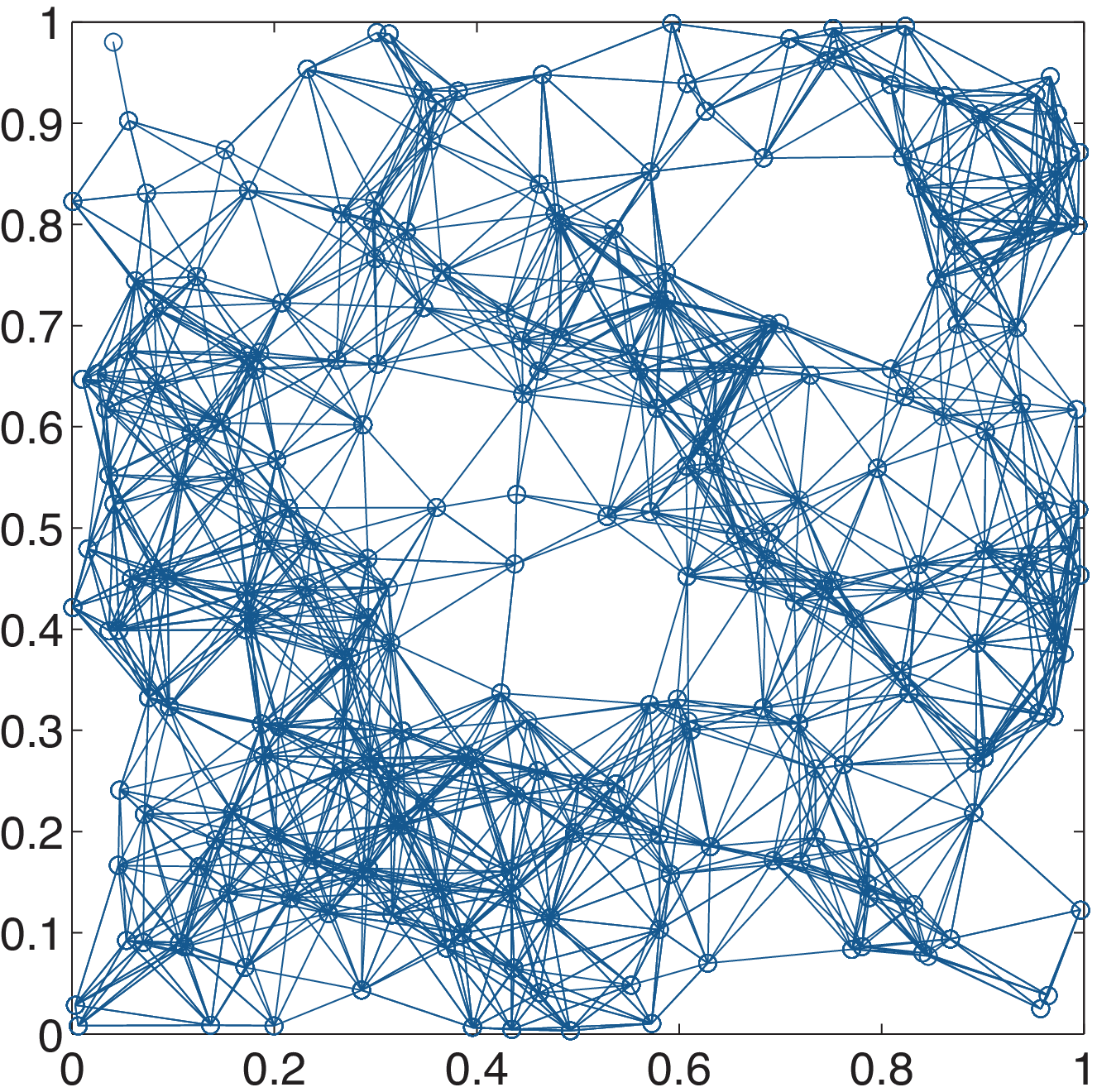} 
\par\end{centering}

\begin{centering}
\includegraphics[width=1\textwidth]{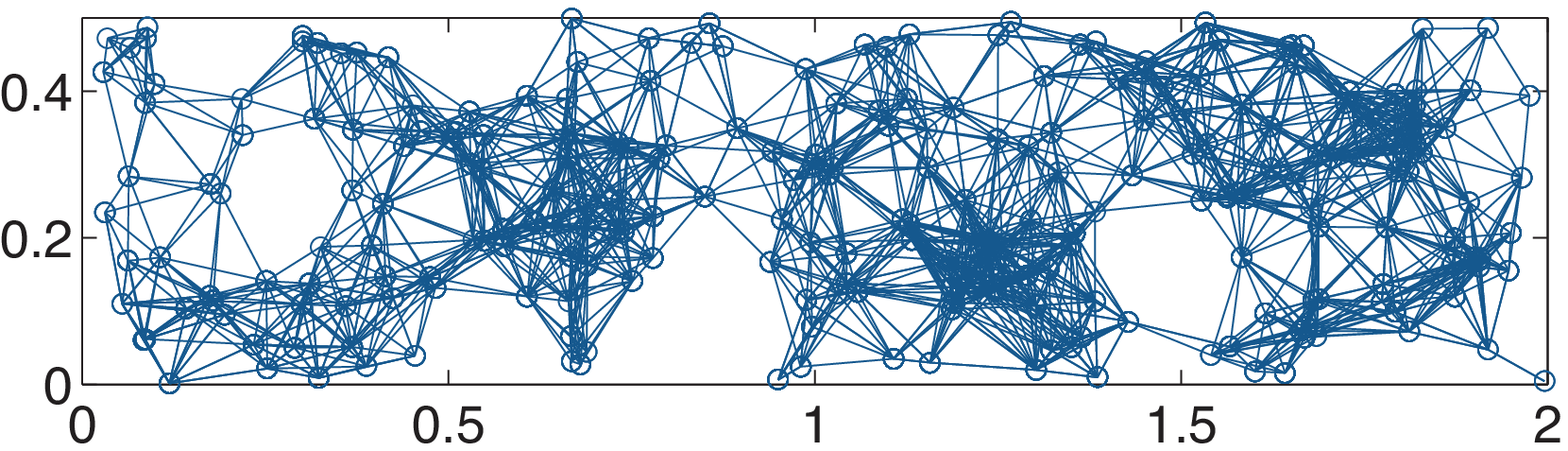} 
\par\end{centering}

\caption{Illustration of a RRG created with 250 nodes embedded into a unit
square, $a=1$, (top) and a unit rectangle with $a=2$ (bottom). In
both cases the nodes are connected if they are at a Euclidean distance
smaller or equal than $r=0.15$.}

\label{RRGs} 
\end{figure}

In Fig.~\ref{RRGs} we illustrate two RRGs with different values
of the rectangle side length $a$ and the same number of nodes and
edges. In the first case when $a=1$ the graph corresponds to the
classical random geometric graph in which the nodes are embedded into
a unit square. The second case corresponds to $a=2$ and it represents
a slightly elongated rectangle.

\subsection{About the connectivity of RRGs}

An important question when studying RGGs in general is related to
the connectivity of the resulting graphs. That is, for which values
of the connection radius is an RGG with $n$ nodes connected with
high probability? In the case of the square, Penrose \citep{Penrose connectivity}
proved that for the two-dimensional case

\begin{equation}
\lim_{n\rightarrow\infty}P\left[\bar{k}-\log n\leq\alpha\right]=\exp\left(-\exp\left(-\alpha\right)\right),\label{eq:ProbConnectivity}
\end{equation}

where $P\left[\cdots\right]$ represents the probability that $\left[\cdots\right]$
takes place and $\bar{k}$ is the average degree.

This means that for $\alpha\rightarrow+\infty$ the RGG is almost
surely connected when $n\rightarrow\infty$, and almost surely disconnected
when $\alpha\rightarrow-\infty$.

In the case of the RRG where the value of $\bar{k}$ depends on the
relation between the two sides of the rectangle we can write (\ref{eq:ProbConnectivity})
for the unit rectangle as \citep{Estrada_Sheerin_1}

\begin{equation}
\lim_{n\rightarrow\infty}P\left[(n-1)f-\log n\leq\alpha\right]=\exp\left(-\exp\left(-\alpha\right)\right),\label{eq:ConnecRRG}
\end{equation}

where $f$ is given by

\begin{equation}
f=\begin{cases}
0\leq r\leq a^{-1} & \pi r^{2}-\frac{4}{3}(a+a^{-1})r^{3}+\frac{1}{2}r^{4},\\
a^{-1}\leq r\leq a & -\frac{4}{3}ar^{3}-r^{2}a^{-2}+\frac{1}{6}a^{-4}+(\frac{4}{3}r^{2}+\frac{2}{3}a^{-2})\sqrt{a^{2}r^{2}-1}\\
 & \quad+2r^{2}\arcsin(\frac{1}{ar}),\\
a\leq r\leq\sqrt{a^{2}+a^{-2}} & -r^{2}(a^{2}+a^{-2})+\frac{1}{6}(a^{4}+a^{-4})-\frac{1}{2}r^{4}\\
 & \quad+(\frac{4}{3}r^{2}a^{-1}+\frac{2}{3}a)\sqrt{r^{2}-a^{2}}+(\frac{4}{3}r^{2}+\frac{2}{3}a^{-2})\sqrt{a^{2}r^{2}-1}\\
 & \quad-2r^{2}(\arccos(\frac{1}{ar})-\arcsin(\frac{a}{r})).
\end{cases}\label{eq:expected values}
\end{equation}

The significance of the function $f$ is clearer when we consider
\cite{Estrada_Sheerin_1} that the expected average degree in a RRG
is given by 
\begin{equation}
\bar{k}=(n-1)f.\label{eq:expected degree}
\end{equation}

Because the parameter $\alpha$ is unknown and it depends on the specific
RRG considered, we have obtained a lower bound for $\exp\left(-\exp\left(-\alpha\right)\right)$
using (\ref{eq:ConnecRRG}):

\begin{equation}
\exp\left(-\exp\left(-\left((n-1)f-\log n\right)\right)\right)\leq\exp\left(-\exp\left(-\alpha\right)\right).\label{eq:lowerbound connectivity}
\end{equation}

\begin{figure}[h]
\begin{centering}
\includegraphics[width=0.4\textwidth]{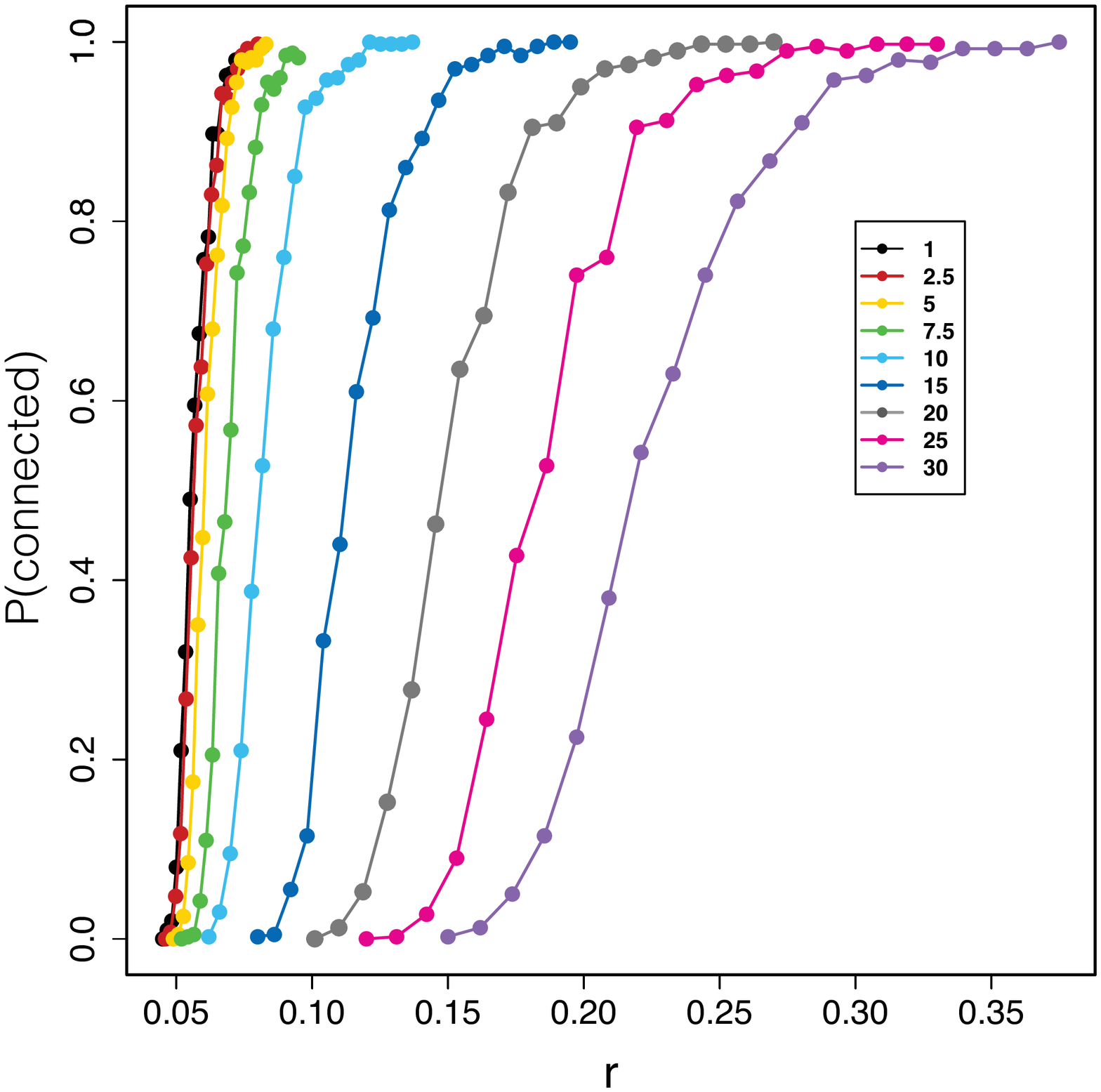}\includegraphics[width=0.4\textwidth]{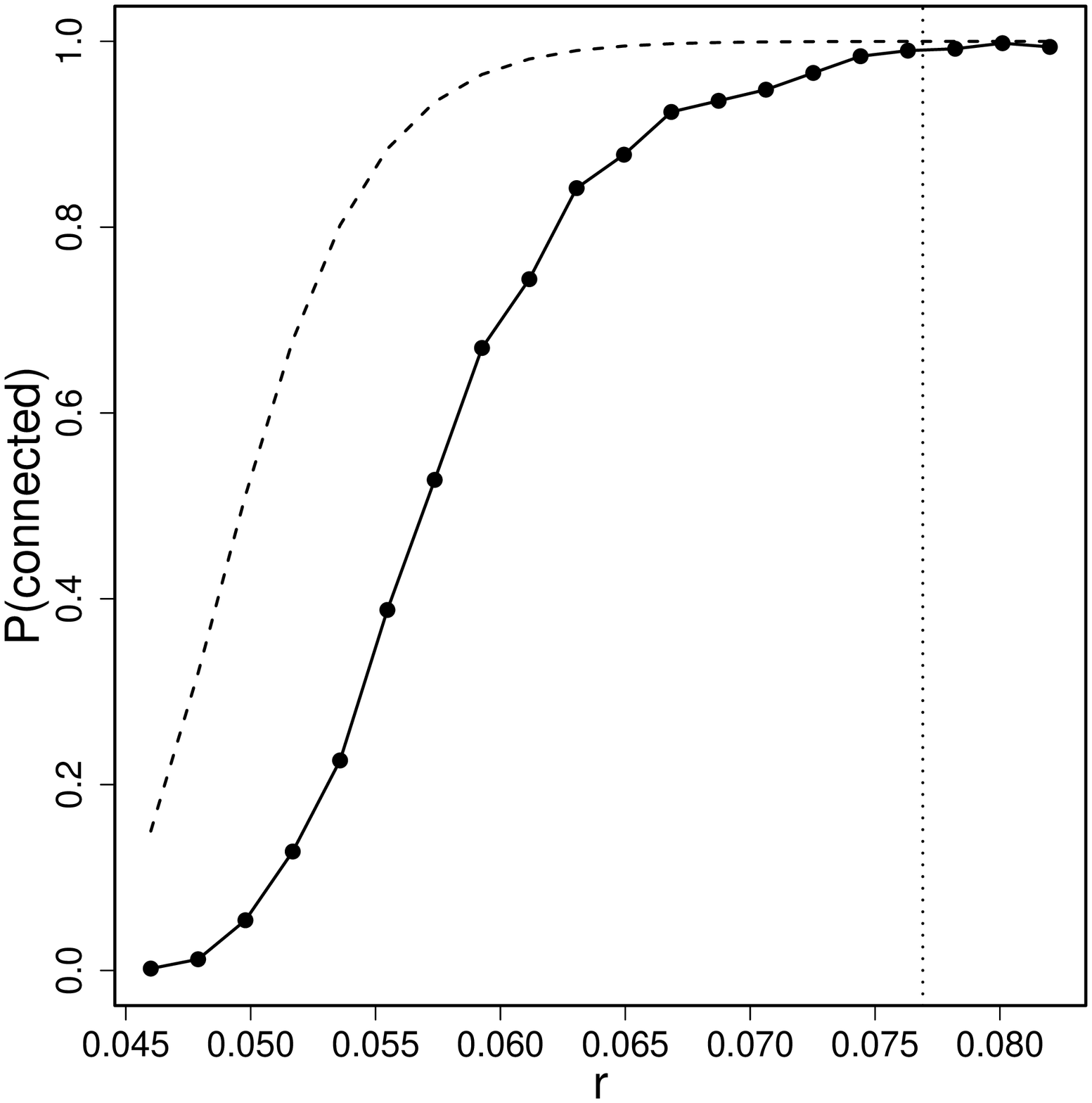}
\par\end{centering}

\begin{centering}
\includegraphics[width=0.5\textwidth]{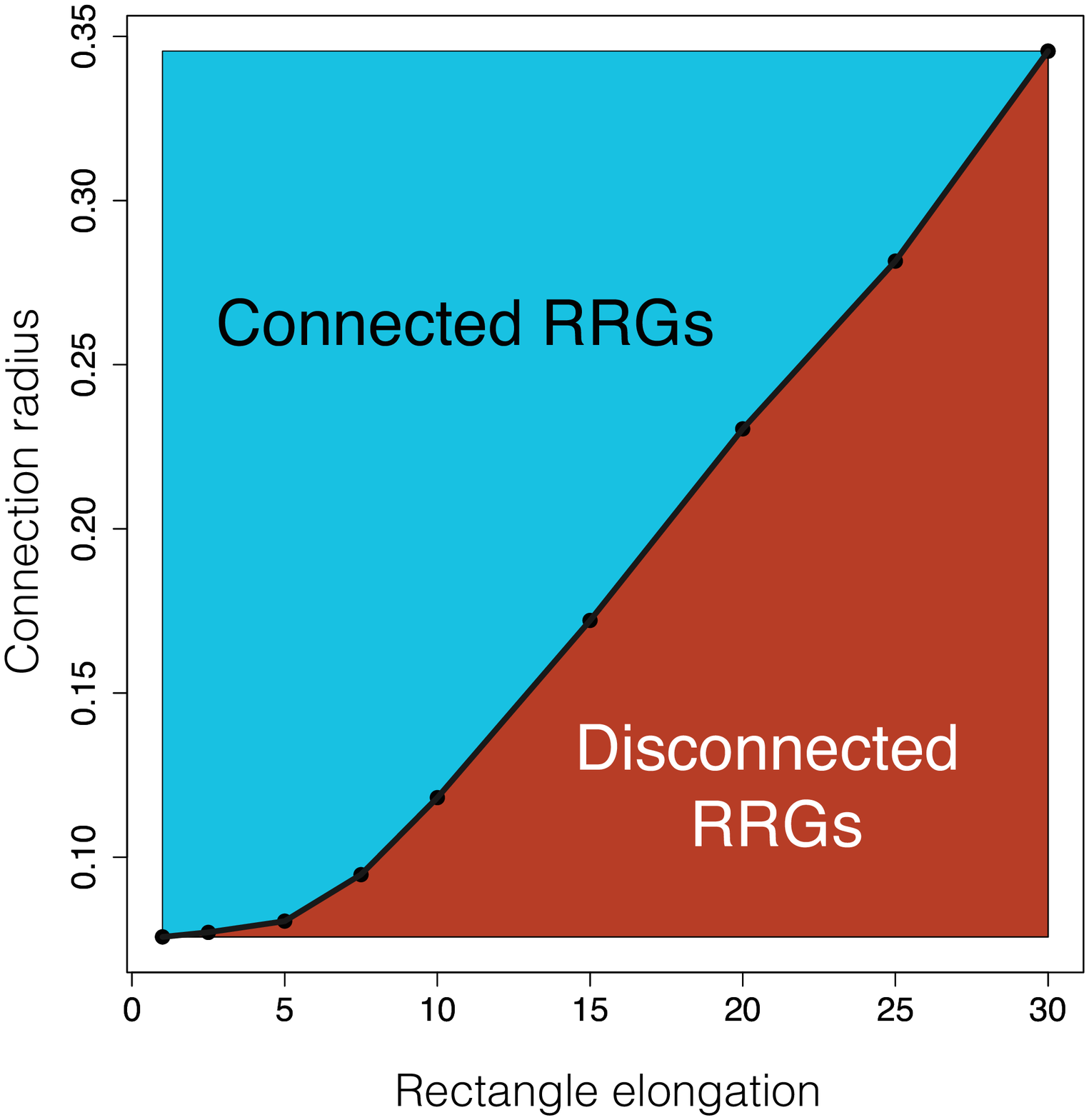} 
\par\end{centering}

\caption{(panel a) Change of the connectivity of RRGs with the change of the
connection radius for different values of the rectangle elongation.
(Panel b) Illustration of the way in which the critical radius for
an RRG is obtained. Also the upper bound (\ref{eq:lowerbound connectivity})
(red dotted line) is illustrated. (Panel c) Plot of the critical versus
the rectangle elongation for the RRGs. The line dividing the two regions
represents the critical values of radius and elongation. All RRGs
studied here have $n=1000$ nodes and all the calculations are the
result of averaging 20 random realizations of the RRG with the given
parameters.}

\label{connectivity} 
\end{figure}

In Fig.~\ref{connectivity} (a) we illustrate the variation of the
connectivity of RRGs with the change of the connection radius for
different values of the rectangle elongation obtained by computational
realizations of the RRGs. As can be seen the probability that the
RRG is connected changes as a sigmoid function with the increase of
the connection radius in a similar way as in the case of the RGGs.
However, as the elongation of the rectangle increases (increase of
$a$) it is more difficult for the graph to be connected and the critical
radius guaranteeing that the graph is connected increases significantly
with $a$. In panel (b) of Fig.~\ref{connectivity} we illustrate
the way in which we determine these critical radii. For a given value
of $a$ we find the minimum value of $r$ for which $P\left(connected\right)=1$.
Although we use in all cases the values obtained from the simulations
we can see that the theoretical bound for $P\left(connected\right)$
(\ref{eq:lowerbound connectivity}) produce very similar results. 

We then plot the values of the connectivity radius versus the elongation
of the rectangles (see Fig.\ref{connectivity} (c)). The curve joining
the points of this plot makes a separation between the RRGs which
are connected (upper triangular part) from those which are disconnected
(lower triangular part). That is, the curve represents the critical
radii versus critical elongation, and it gives the critical region
indicating the connectivity of the RRGs. It can be read in two different
ways. You can fix a value of $a$ and then determine which is the
critical radius for which the network will be disconnected. For instance,
for a rectangle with longer side $a=15$ it is necessary to use radius
larger than 0.17 to make the RRGs connected. More interesting for
this work is the other way around. That is, we have a fixed radius
of connection, which may represent the radius of spreading of a disease
among plants. Then, you can find what elongation of the rectangle
disconnects the network. For instance, if the connection radius is
fixed to $r=0.35$ every RRG is connected for $a<30$. Then, we emphasize
here that we always work in the region of connected RRGs in this work.
That is, any analysis carried out in this paper is based on graphs
for which the connectivity of the graph is 100\% guaranteed as we
work in the upper triangular part of this plot. In addition, we check
computationally that every RRG generated in this work is connected.

\section{Epidemics on Networks}

The spreading of an infectious disease on networks can be modeled
representing individuals as nodes and the contacts between them as
edges. In this context individuals are categorized in different compartments
according to their health state \citep{Epidemics_book1}: i.e.~\textit{susceptible}
(S) for individuals that can be infected by the disease, \textit{infected}
(I) for infectious individuals that can spread the pathogen or \textit{recovered}
(R) for individuals that already passed the disease and are immune
to it.

Two fundamental models for disease spreading are the so-called SIS
and SIR \citep{Epidemics_book1,Epidemics_book2}. The SIS is intended
to model recurrent diseases that do not provide immunity, i.e.~the
common cold or most sexually transmitted diseases, where individuals
can get the infection multiple times during their lifetime. Instead,
in the SIR, once an individual get cured from the disease she enters
the recovered compartment and cannot be infected again, that is, she
acquires immunity. Both SIS and SIR dynamics are governed by two parameters,
namely: the per contact \textit{infection rate} $\beta$ and the \textit{recovery
rate} $\mu$. Let , $s_{i}$, $x_{i}$ and $r_{i}$ be the probabilities
that the node $i$ is susceptible, infected or has recovered from
infection, respectively. The equations governing a SIS process are
the following:

\begin{align}
\dot{s}_{i} & =-\beta s_{i}\sum_{j}A_{ij}x_{j}+\mu x_{i},\\
\dot{x}_{i} & =\beta s_{i}\sum_{j}A_{ij}x_{j}-\mu x_{i},
\end{align}

while those governing a SIR one are:

\begin{align}
\dot{s}_{i} & =-\beta s_{i}\sum_{j}A_{ij}x_{j},\\
\dot{x}_{i} & =\beta s_{i}\sum_{j}A_{ij}x_{j}-\mu x_{i},\\
\dot{r}_{i} & =\mu x_{i}.
\end{align}

In these models, $\beta$ represents the rate for a susceptible individual
to catch the disease once in contact with an infected one through
a link of the network, whereas the recovery rate $\mu$ characterizes
the rate at which an infected recovers from the disease. Note that
the spreading of the disease depends on the network of contacts (an
isolated individual can not catch the disease), while the recovery
phase is independent of the substrate network (an isolated infected
individual will recover after some time).

The ratio $\beta/\mu$ drives the spreading of the disease. Depending
on its infectious power two distinct phases are possible: an absorbing
one where the spreading is not efficient enough to reach a large fraction
of the system and the disease is absorbed and an active phase where
the epidemics reaches a macroscopic fraction of the network. The transition
from the absorbing to the active phase strictly resembles a non-equilibrium
second order phase transition in statistical physics \citep{Transitions_book1,Transitions_book2}.
The critical value of this transition $\left(\frac{\beta}{\mu}\right)_{c}=\tau$
is defined as the \textit{epidemic threshold}. This term is also known
as the\textit{ basic reproduction number} and it represents a threshold
in the sense that when $\tau<1$ the infection dies out and if $\tau>1$
the disease becomes an epidemic. In those cases where $\tau=1$, the
disease remains in the population becoming endemic. The value of this
threshold strongly depends on the topology of the network. In particular,
for a given graph $G=\left(V,E\right)$, it has been shown that \citep{Epidemics_2,Epidemics_4,Epidemics_5}:

\begin{equation}
\tau=\dfrac{1}{\lambda_{1}\left(G\right)},\label{eq:epidemic threshold}
\end{equation}

where $\lambda_{1}\left(G\right)$ is the largest eigenvalue of the
adjacency matrix of the network. In the case of RGG, Preciado and
Jadbabaie \citep{Preciado} have made an exhaustive spectral analysis
of virus spreading using the spectral moments of the adjacency matrix.
They have found that for the RGG in $d$-dimensional cube, the spectral
radius is bounded as $\lambda_{1}\left(G\right)<c_{d}nr^{d}$, where
$c_{d}$ is a constant characteristic of the RGG in the $d$-dimensional
cube. In 2-dimensional space this means that $\lambda_{1}\left(G\right)<c_{2}nr^{2}$.
Then, because in these graphs the average degree is $nr^{2}$, the
previous expression basically tells us that the spectral radius is
bounded by the average degree of the RGG.

\section{Epidemic Threshold in RRGs}

In this section we concentrate more on the phenomenology of the process
than in deriving analytical results about the dependance of the epidemic
threshold with the topological parameters of the RRGs. Thus, we will
obtain some sort of mean-field approach that captures the behavior
of the disease parameters with the topological ones. We start by considering
the following well-known bounds for the largest eigenvalue of the
adjacency matrix of a simple graph $\lambda_{1}\left(G\right)=\lambda_{1}$ 

\begin{equation}
\bar{k}\leq\lambda_{1}\leq k_{max},
\end{equation}

where $k_{max}$ and $\bar{k}$ are the maximum and the average degree,
respectively. Then, it is straightforward to realize that

\begin{equation}
\tau\geq\dfrac{1}{\bar{k}}.\label{eq:thre}
\end{equation}

Then, by replacing (\ref{eq:expected degree}) into (\ref{eq:thre})
we have the following bound for the epidemic threshold of a RRG

\begin{equation}
\tau\geq\dfrac{1}{\left(n-1\right)f}.\label{eq:thre-1}
\end{equation}

This result generalizes the one obtained by Preciado and Jadbabaie
\citep{Preciado} for the RGG to the case where we have a rectangle
of any elongation and where we consider explicitly the border effects
of the rectangle (respectively the square in RGG).

\begin{figure}
\begin{centering}
\includegraphics[width=0.33\textwidth]{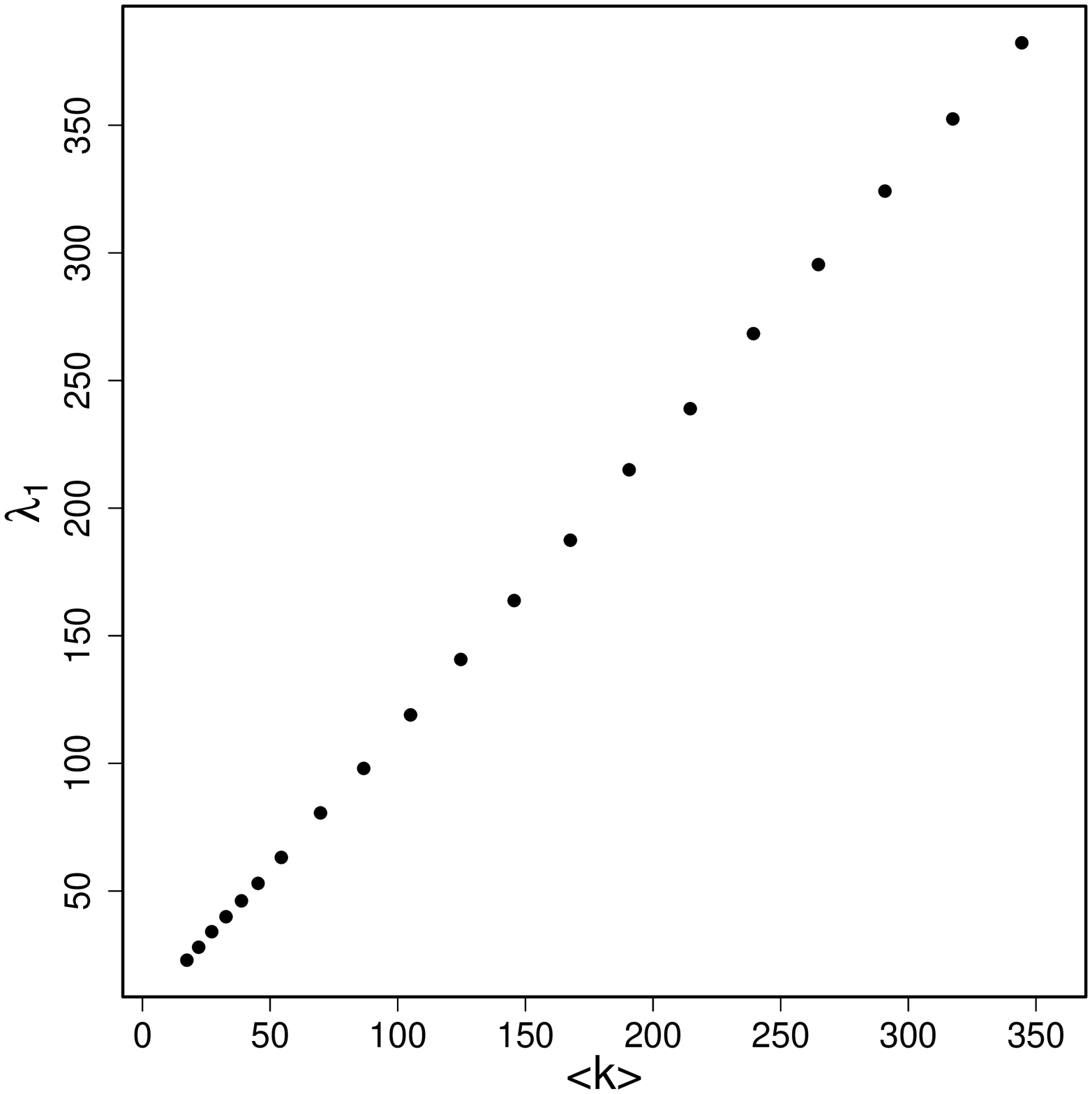}\includegraphics[width=0.33\textwidth]{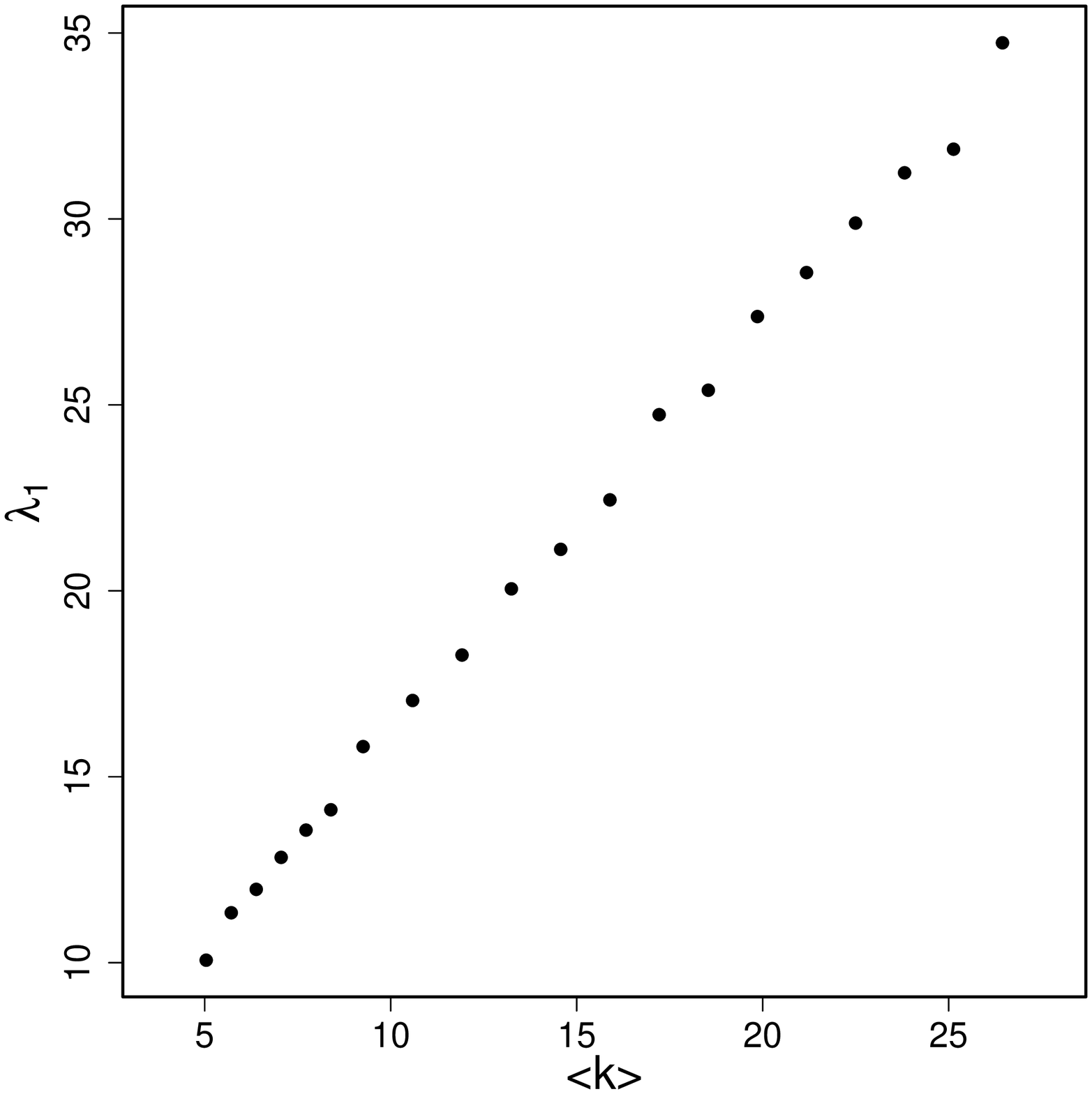}\includegraphics[width=0.33\textwidth]{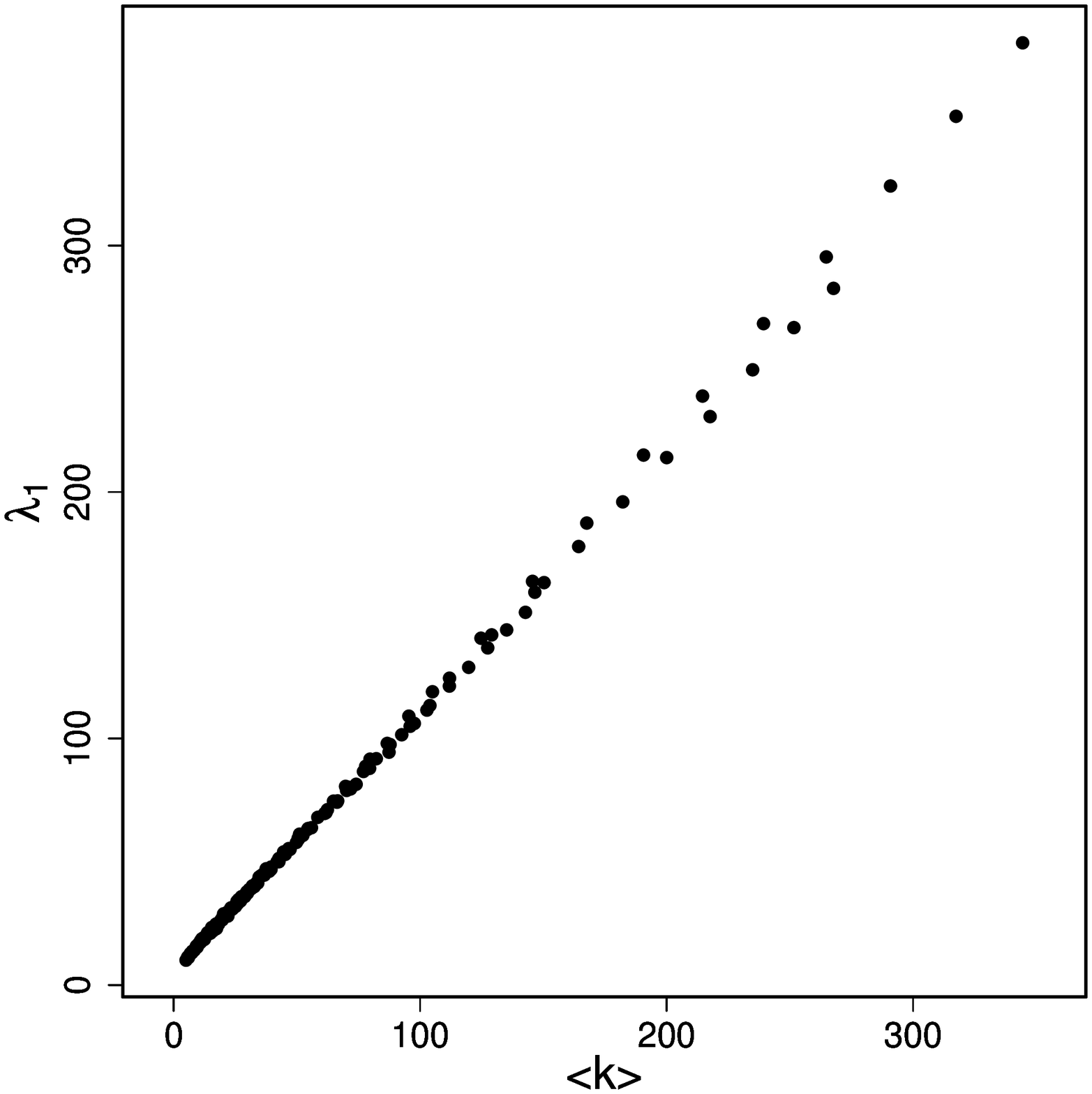}
\par\end{centering}

\caption{Scatter plots of the spectral radius versus the average degree for
RRGs with $a=1$ (left), $a=30$ (centre) and $a=1,2.5,5,7.7,15,20,25,30$
(right) for different values of the connection radius.}

\label{Spectral radius vs k_mean}

\end{figure}

Let us now consider what happen to the epidemic threshold when we
elongate the rectangle without disconnecting the RRG. That is, what
is the behavior of the epidemic threshold when $a\rightarrow a_{c}$.
In the following we will prove that when $a\rightarrow a_{c}$, $f$
decreases. Consequently, the spectral radius of the adjacency matrix
$\lambda_{1}$ also decreases when $a\rightarrow a_{c}$, which implies
that the epidemic threshold grows monotonically. Our strategy here
is to prove that the increase of the rectangle elongation produces
a decay of the average degree of the RRGs. Strictly speaking, proving
that the average degree decreases with the elongation of the rectangle
is not a prove that the spectral radius also decreases. Although such
a proof is analytically possible, it is out of our intention of working
more on the phenomenology of the process in this work. However, it
is possible to indirectly infer such relation between the elongation
and the spectral radius as follow. First, it is well-known in spectral
graph theory that the average degree is a tight bound to the spectral
radius of graphs in general. In particular, we are interested here
in showing whether the average degree and the spectral radius of RRGs
show the same behavior when the rectangle is elongated. In Figure
\ref{Spectral radius vs k_mean} we illustrate the the plot of the
spectral radius versus the average degree for RRGs with $a=1$ (left),
$a=30$ (centre) and $a=1,2.5,5,7.7,15,20,25,30$ (right) for different
values of the connection radius. As can be seen in all cases, and
particularly in the last one, the trend of the spectral radius and
the average degree is exactly the same and indeed they are very highly
linearly correlated. Thus, our conclusion here is that proving that
the average degree have certain behavior when the rectangle is elongated
can be directly extrapolated to the behavior of the spectral radius
with the elongation of the rectangle.

Then, in order to prove this result we first consider what happens
to the function $f$ when $a\rightarrow a_{c}$. Let $0<r\leq a^{-1}$.
Then, the first derivative of $f_{1}=f\left(0\leq r\leq a^{-1}\right)$
is given by

\begin{equation}
\dfrac{\partial f_{1}}{\partial a}=-\dfrac{4}{3}r^{3}\left(1-a^{-2}\right),
\end{equation}

and since

\[
\left(1-a^{-2}\right)\geq0.
\]

this is bounded by

\[
\dfrac{\partial f_{1}}{\partial a}\leq0.
\]

Let $a^{-1}\leq r\leq a$. Then, the first derivative of $f_{2}=f\left(a^{-1}\leq r\leq a\right)$
is given by

\begin{eqnarray}
\dfrac{\partial f_{2}}{\partial a} & = & \dfrac{-4r^{3}}{3}+\dfrac{2r^{2}}{a^{3}}-\dfrac{2}{3a^{5}}+\dfrac{4(a^{4}r^{4}-2a^{2}r^{2}+1)}{3a^{3}\sqrt{a^{2}r^{2}-1}}
\end{eqnarray}

which is bounded as,

\begin{equation}
\dfrac{\partial f_{2}}{\partial a}\leq h<0,
\end{equation}

where

\begin{eqnarray}
h & = & \lim_{r\rightarrow a}\dfrac{\partial f_{2}}{\partial a}=\dfrac{2}{a}-\dfrac{4a^{3}}{3}-\dfrac{2}{3a^{3}}+\dfrac{4(a^{8}-2a^{4}+1)}{3a^{3}\sqrt{a^{4}-1}}.
\end{eqnarray}

Let $a\leq r\leq\sqrt{a^{2}+a^{-2}}$. Then, the first derivative
of $f_{3}=f\left(a\leq r\leq\sqrt{a^{2}+a^{-2}}\right)$ is given
by

\begin{eqnarray}
\dfrac{\partial f_{3}}{\partial a} & =2r^{2}\left(\dfrac{1}{a^{3}}-a\right)+\dfrac{2}{3}\left(a^{3}-\dfrac{1}{a^{5}}\right)+\dfrac{4(a^{4}r^{4}-2a^{2}r^{2}+1)}{3a^{3}\sqrt{a^{2}r^{2}-1}}-\dfrac{4(a^{4}-2a^{2}r^{2}+r^{4})}{3a^{2}\sqrt{r^{2}-a^{2}}} & ,
\end{eqnarray}

which is bounded as,

\begin{equation}
\dfrac{\partial f_{2}}{\partial a}\leq g<0,
\end{equation}

where

\begin{eqnarray}
g & = & \lim_{r\rightarrow t}\dfrac{\partial f_{3}}{\partial a}=0.
\end{eqnarray}

and $t=\sqrt{a^{2}+a^{-2}}$. Then, because all the derivatives are
negative, we have proven the result. Strictly speaking the fact that
$1/\left(n-1\right)f$ increases with increasing $a$ does not necessarily
imply that $\tau$ will exhibit a similar trend for every $a$. However,
as we will see in the next section there is a very good linear correlation
between the values of $\tau$ obtained from the simulations and the
lower bound $1/\left(n-1\right)f$ , which indicates that both quantities
follow the same trend as consequently that the previous assertion
relating the behavior of the epidemic threshold when when $a\rightarrow a_{c}$
is general. We discuss this in more detail in the net section.

\begin{figure}[h]
\begin{centering}
\includegraphics[width=0.75\textwidth]{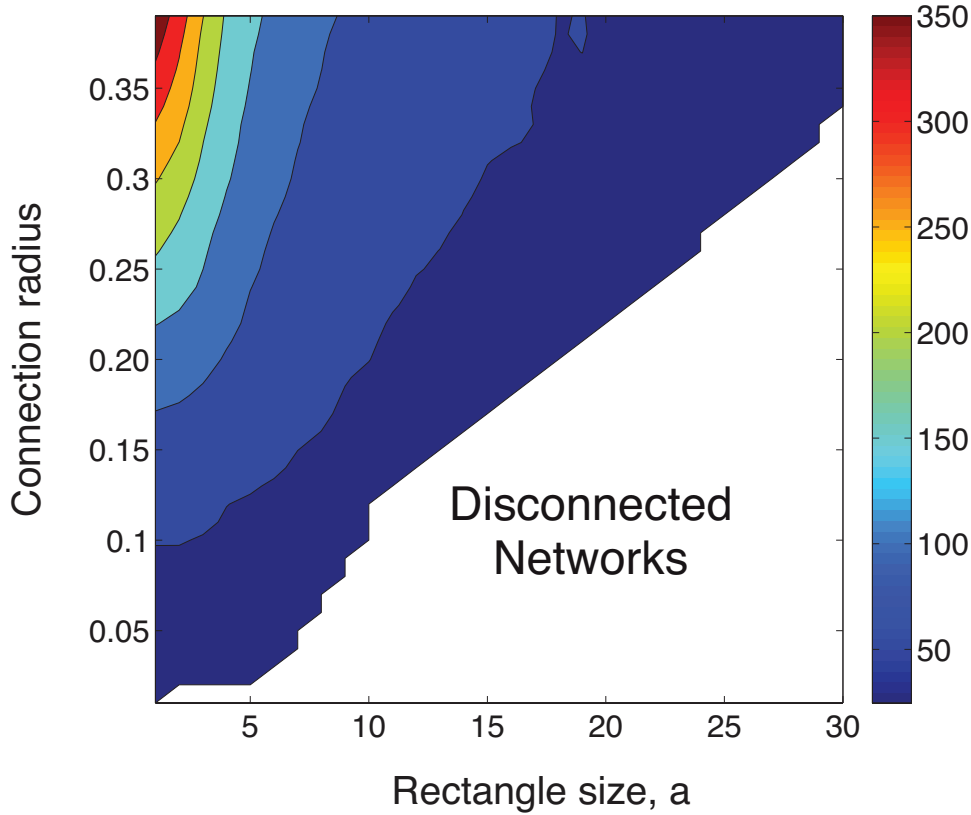} 
\par\end{centering}

\caption{Values of the spectral radius $\lambda_{1}$ of the adjacency matrix
of RRGs with $n=1000$ nodes as a function of the rectangle size length
$a$ and the connection radius $r$. The bottom-right part of the
plot corresponds to networks which are created with radius below the
critical radius, $r<r_{c}$ (see plot (c) in Fig.~\ref{connectivity}),
and consequently are disconnected. All the calculations are the result
of averaging 20 random generations of the RRG with the given parameters. }

\label{spectral radii} 
\end{figure}

In the Fig.~\ref{spectral radii} we illustrate the plot of the spectral
radius of the adjacency matrix of RRGs as a function of both the rectangle
size length $a$ and the connection radius $r$. As expected the lower
triangular part of the plot corresponds to the disconnected RRGs,
which are never used in this work. However, in the upper triangular
part of the plot we observe a large variation of the spectral radius
$\lambda_{1}$ of an RRG with both parameters of the model. For a
fixed connection radius the values of $\lambda_{1}$ decay with the
elongation of the rectangle as expected from the previous analytical
results. Notice that $\lambda_{1}$ can change as much as from $380$
to $35$ for a constant radius $r=0.4$ when changing the rectangle
size from $a=1$ to $a=30$. This, of course, is the main cause of
the change of the epidemic threshold predicted by the bound (\ref{eq:thre})
obtained at the beginning of this section.

\section{Epidemics on RRGs. Simulations}

In this section we conduct extensive numerical simulations of the
SIS dynamics for different values of the elongation $a$ and fixed
radius $r$ with the goal of checking the goodness of the bound defined
in Eq.~\ref{eq:thre} and to illustrate how the elongation of the
rectangle in the RRG model changes the epidemic dynamics. In the simulations
we start seeding the infection in a small fraction $\rho_{0}=0.01$
of the nodes and let the SIS dynamics evolve for $5\cdot10^{4}$ time-steps.
At this point, we let the simulations run for an additional $10^{3}$
time-steps and calculate the fraction of infected nodes $\rho$ as
the average of $\rho(t)$ over this period. For each selection of
the parameters we performed $250$ independent runs with different
initial conditions. The final value of $\rho$ is obtained as the
average over all the runs.

\begin{figure}[h]
\begin{centering}
\includegraphics[width=0.75\textwidth]{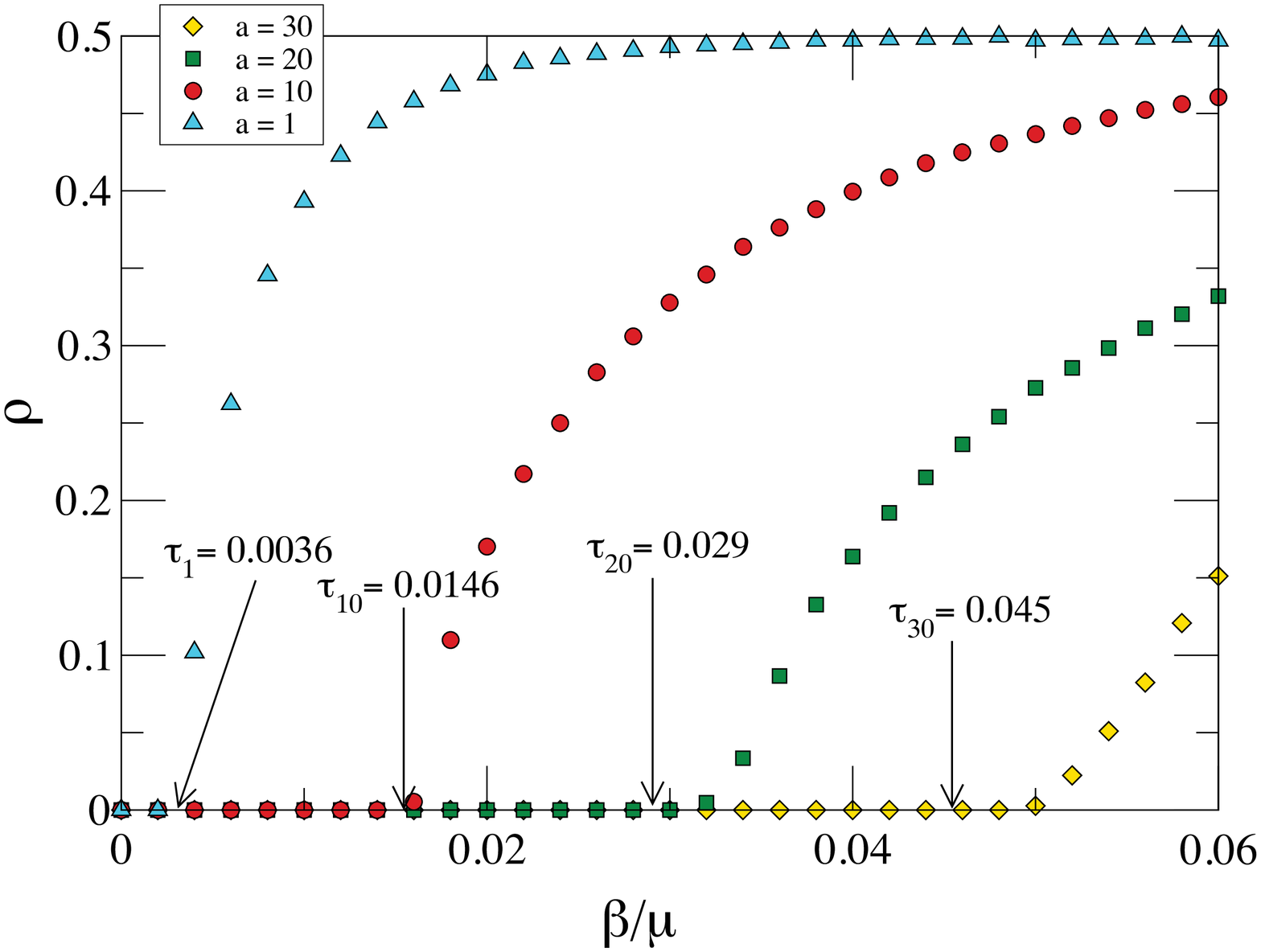} 
\par\end{centering}

\centering{}\caption{(a) Fraction of infected nodes at the stationary state $\rho$ as
a function of the infection rate $\beta$ for different values of
$a=1,10,20,30$. $a=1$ represents the first case ($0\leq r\leq a^{-1}$)
of Eq.~\ref{eq:expected values} while $a=10,20,30$ fall in the
second case ($a^{-1}\leq r\leq a$). Other parameters are: $n=10^{3}$
nodes, $r=0.35$ and $\mu=1.0$. Each point is an average over $250$
independent runs. The values shown by arrow are the analytical ones
obtained using (\ref{eq:thre-1}).}
\label{stationary} 
\end{figure}

Figure~\ref{stationary} shows the fraction of infected nodes in
the stationary state against the infection rate $\beta$ for different
values of $a=1,10,20,30$. The values shown by arrow are the analytical
ones obtained using (\ref{eq:thre-1}).

\begin{figure}[h]
\centering{}\includegraphics[width=0.49\textwidth]{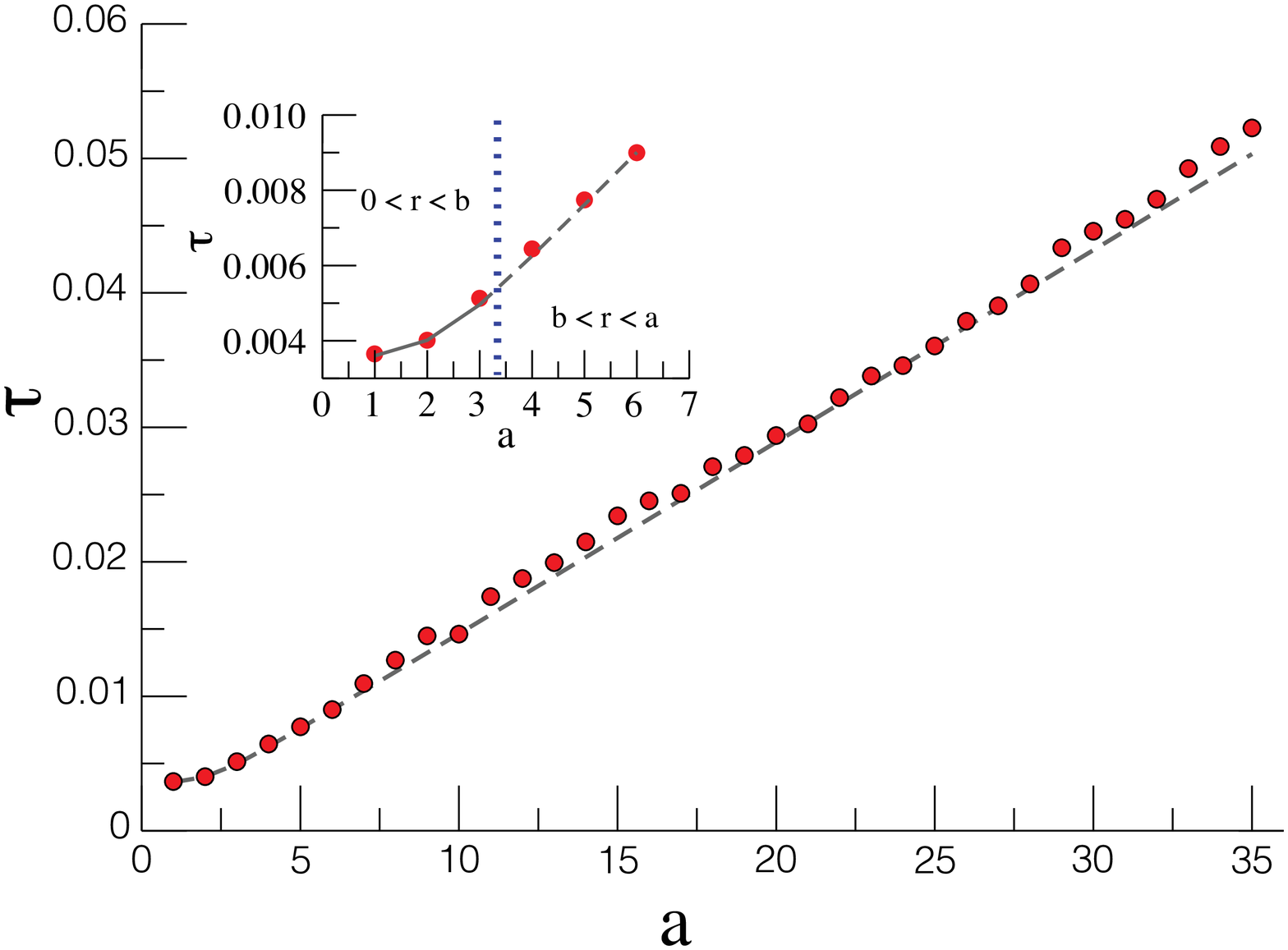}
\includegraphics[width=0.49\textwidth]{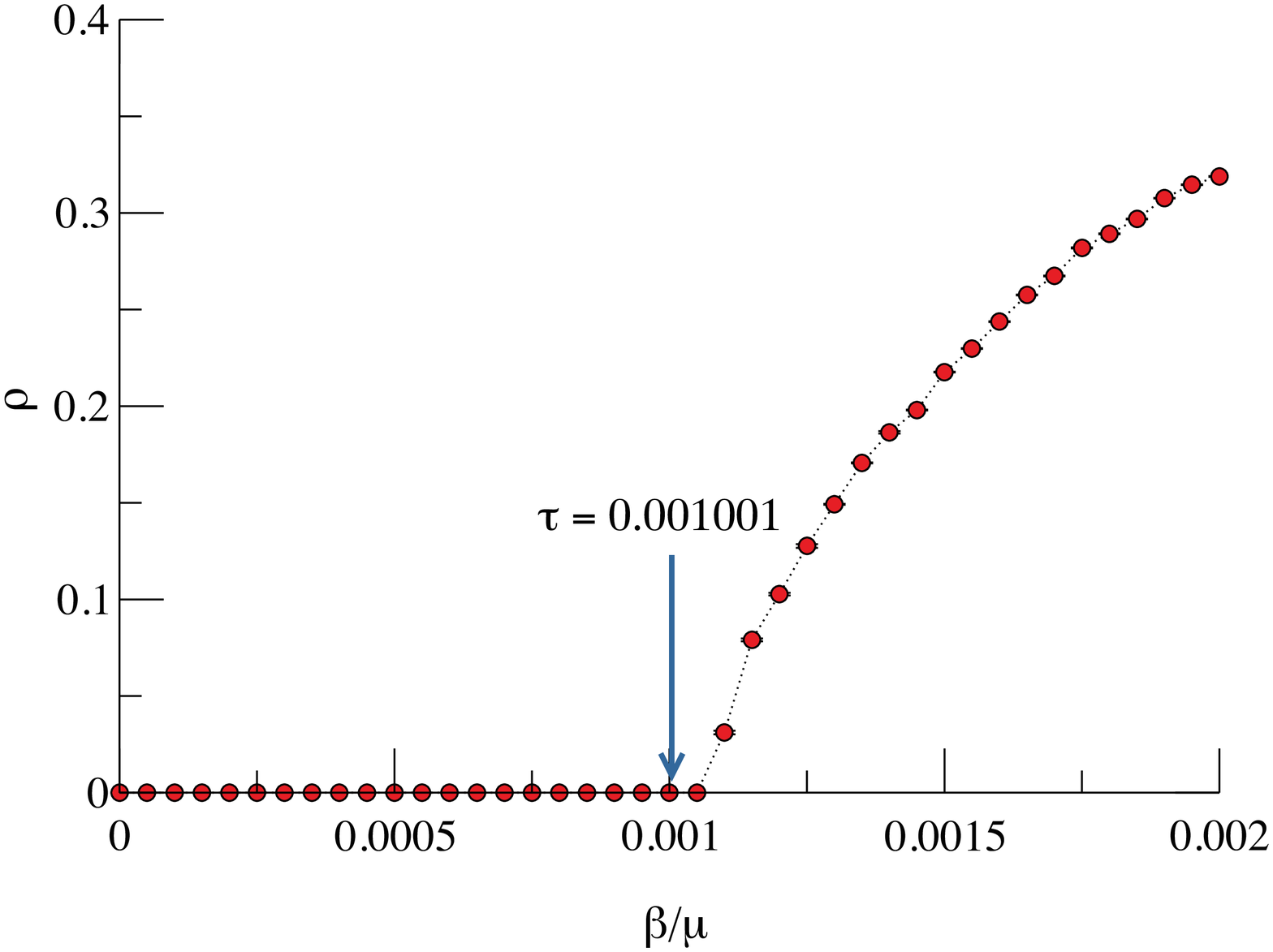} \caption{(panel a) Comparison between the theoretical bound and the epidemic
threshold obtained via numerical simulations. Line represents the
theoretical prediction of Eq.~\ref{eq:thre} while points represent
the numerical threshold. The inset shows a zoom for the first case
of Eq.~\ref{eq:expected values} $0\leq r\leq a^{-1}$ (full line)
and the second case $a^{-1}\leq r\leq a$ (dashed line). Other parameters
are: $n=10^{3}$ nodes, $r=0.35$ and $\mu=1.0$. Each point is average
over $250$ independent runs. (panel b) Fraction of infected nodes
at the steady state $\rho$ as a function of the infection rate $\beta$
for $a\leq r\leq\sqrt{a^{2}+a^{-2}}$. In the simulations $a=3$ and
$r=3.01$. Other parameters are: $n=10^{3}$ nodes and $\mu=1.0$.
Each point is average over $250$ independent runs.}
\label{threshold} 
\end{figure}

To have a more detailed picture of the behavior of the epidemic threshold,
in Fig.~(\ref{threshold}a) we compare the theoretical bound with
the epidemic threshold obtained via the numerical simulations. As
we have stressed in the previous section this comparison is very important
for understanding whether the epidemic threshold and the bound $1/\left(n-1\right)f$
follow the same trend with the elongation of the rectangle. Our comparison
covers two of the three cases of Eq.~\ref{eq:expected values}: $0\leq r\leq a^{-1}$
and $a^{-1}\leq r\leq a$ respectively. As can be seen in this Figure
the lower bound \ref{eq:thre-1} is very tight, and more importantly
the bound and the 'observed' epidemic threshold display the same behavior
when the rectangle elongation change. Indeed, our analysis of the
difference between the observed value of the epidemic threshold and
the lower bound obtained by Eq.~\ref{eq:thre}  shows that for all
the RRGs having $1\leq a\leq35$ such relative difference is 2.93\%
and in no case it is larger than 10\%. Also we observe no trend in
the relative difference related to the elongation of the rectangle.
That is, the relative difference is neither increasing nor decreasing
with the elongation of the rectangle.

Finally, in Fig.~\ref{threshold}b we also tested the third case
of Eq.~\ref{eq:expected values}, $a\leq r\leq\sqrt{a^{2}+a^{-2}}$
for $a=3$ and $r=3.01$. In all cases, as expected, the theoretical
and the simulation results show that the increase of the elongation
of the rectangle produces an increase of the epidemic threshold. In
other words, the elongation of the rectangle retards the disease progress
through the nodes embedded in the rectangle.

In the case of disease propagating on plants, these results---both
analytical and simulations---coincide with the field observations
and simulations using stochastic models \citep{square plots,square fields_1,plot shapes,square fields_2,rectangular field coconut,plot size_1,computer simulations_1,stochastic model,mathematical model}
which suggest that square plots and fields favored higher spreading
of plant diseases than elongated ones of the same area \citep{square plots,square fields_1,plot shapes,square fields_2}.

Our analytical and simulation results point to the fact that under
the same conditions, the propagation of an epidemic on a rectangular
plot/field is much harder than on a square one because a larger number
of infected individuals is needed for the disease to become epidemic.
Here we have kept the size of the plot/field constant by considering
unit rectangles in our analysis. However, it is important to consider
that other factors, such as the orientation of the plot/field play
fundamental role in the propagation of a disease on plants. For instance,
if the rectangular plots are placed perpendicular to the direction
of the prevalent winds the disease will not propagate as a consequence
of this factor.

\section{Conclusions}

We have studied the propagation of diseases on a recently proposed
random rectangular graph (RRG) model, deriving analytically a lower
bound of the epidemic threshold for a SIS or SIR model running on
these networks. This model is appropriate for the simulation of disease
spreading on plants allocated on plots and field of varied shapes.
RRGs account for the spatial distribution of nodes allowing the variation
of the shape of the unit square commonly used in random geometric
graphs (RGGs). We have shown here by using analytical results and
extensive numerical simulations of the SIS dynamics for different
values of the elongation $a$ and a fixed radius $r$ that the elongation
of the plots/fields in which the nodes (plants) are distributed makes
the network more resilient to the propagation of epidemics. This is
due to the fact that the epidemic threshold increases with the elongation
of the rectangle. These results agree with a large accumulation of
empirical evidence about the role of plots/fields elongation on the
propagation of diseases on plants. This model represents a new way
to analyze disease propagation on plants or similar scenarios, by
combining the heterogeneities introduced at individual level by networks
with the influence produced by the shape variation of the plots and
fields where the plants are growing. 
\begin{acknowledgments}
EE thanks the Royal Society of London for a Wolfson Research Merit
Award. SM and YM thank partial support by by MINECO and FEDER funds (grant FIS2014-55867-P);
Comunidad de Arag\'on (Spain) through a grant to the group FENOL; and
the EC Proactive project MULTIPLEX (contract no. 317532). SM is also
supported by MINECO through the Juan de la Cierva Program. MS is supported
by Weir Advanced Research Centre, University of Strathclyde and EPRSC,
UK. \end{acknowledgments}

\end{document}